\documentclass{mem}

\usepackage{natbib}
\usepackage{txfonts}
\usepackage{balance}
\usepackage{graphicx}
\usepackage[a4paper]{hyperref}
\idline{00}{73}

\begin{document}

\title{Peanut-shaped bulges in face-on disk galaxies}

\subtitle{}

\author{J. M\'endez-Abreu\inst{1} \and E. M. Corsini\inst{2}
  \and V. P. Debattista\inst{3} \and S. De Rijcke\inst{4}
  \and J. A. L. \, Aguerri\inst{1} \and A. Pizzella\inst{2} 
}

\offprints{Jairo M\'endez-Abreu; \mbox{\email{jairo@iac.es}}}
 
\institute{Instituto de Astrof\'isica de Canarias,
  La Laguna, Spain
\and
  Dipartimento di Astronomia, Universit\`a di Padova, Padova, Italy
\and
  Centre for Astrophysics, University of Central Lancashire, Preston,
  UK
\and
  Sterrenkundig Observatorium, Universiteit Gent, Gent, Belgium }

\authorrunning{M\'endez-Abreu}

\titlerunning{Peanut-shaped bulges in face-on galaxies}

\abstract{We present high resolution absorption-line spectroscopy of 3 face-on galaxies, NGC~98, NGC~600, and NGC~1703 with the aim of searching for box/peanut (B/P)-shaped bulges.  These observations test and confirm the prediction of \citet{deb_etal_05} that face-on B/P-shaped bulges can be recognized by a double minimum in the profile of the fourth-order Gauss-Hermite moment $h_4$. In NGC~1703, which is an unbarred control galaxy, we found no evidence of a B/P bulge.  In NGC~98, a clear double minimum in $h_4$ is present along the major axis of the bar and before the end of the bar, as predicted. In contrast, in NGC 600, which is also a barred galaxy but lacks a substantial bulge, we do not find a significant B/P shape.
 }

\maketitle{}

\section{Introduction}

Understanding how bulges form is of great importance to develop a
complete picture of galaxy formation.  The processes by which bulges
form are still debated.  On the one hand, the merger of dwarf-sized
galactic subunits has been suggested as the main path for bulge
formation \citep{kau_etal_93}, which is supported by the relatively
homogeneous bulge stellar populations of the Milky Way and M31
\citep{zoc_etal_03}.  Bulges formed in such mergers are termed
'classical' bulges.  Alternatively, bulges may form via internal
`secular' processes such as bar-driven gas inflows, bending
instabilities, clump instabilities, etc.  \citep{com_san_81,
  deb_etal_06}.  Evidence for secular bulge formation includes the
near-exponential bulge light profiles \citep{and_san_94}, a
correlation between bulge and disk scale lengths \citep{mac_etal_03,
  men_etal_08}, the similar colors of bulges and inner disks
\citep{pel_bal_96}, substantial bulge rotation \citep{kor_ken_04}, and
the presence of box/peanut (B/P)-shaped bulges in $\sim45$\% of
edge-on disk galaxies \citep{lut_etal_00}.  A review of secular
`pseudo-bulge' formation and evidence for it can be found in
\citet{kor_ken_04}.  Standard cold dark matter cosmology predicts that
galaxies without classical bulges should be rare \citep{don_bur_04}.
Not only are they not rare in nature, but the formation of
pseudo-bulges means that some fraction of bulged galaxies lack a
classical bulge, exacerbating the disagreement between theory and
observations \citep{deb_etal_06}.  It is therefore important to
determine which bulges are of the classical versus pseudo variety, and
which are a mix of both.

Of primary concern here, several pieces of evidence point to the
identification of most B/P bulges in edge-on spiral galaxies with the
bars of barred spirals.  N-body simulations show that barred
galaxies have a tendency to develop B/P bulges.  The observed
incidence of B/P bulges is consistent with that expected if they are
associated with relatively strong bars.  A recent work by
\citet{lut_etal_00} demonstrated  that $45$\% of 
all bulges are B/P, while amongst those the shape of the bulge
depends mainly on the viewing angle to the bar.  As shown by the
numerical simulations, true peanuts are bars seen side-on, that is,
with the major-axis of the bar perpendicular to the line of sight.
For less-favourable viewing angles, the bulge/bar looks boxy, and if
the bar is seen end-on it looks almost spherical.

The presence of bars in edge-on galaxies with B/P bulges has been
observationally established by the kinematics of gas and stars
\citep[see][and  references therein]{bur_ath_05}.   The  kinematics of
discs harbouring a B/P bulge, as measured from both ionized-gas
emission lines and stellar absorption lines, show the behaviour
expected of barred spirals viewed edge-on. However, the degeneracy
inherent in deprojecting edge-on galaxies makes it difficult to study
other properties of the host galaxy.  For example, simulations show
that a bar can produce a B/P shape even if a massive classical bulge
formed before the disk \citep{deb_etal_05}.
Understanding the relative importance of classical and pseudo-bulges
requires an attempt at a cleaner separation of bulges, bars and
peanuts, which is easiest to accomplish in less inclined systems.

Recently, \citet{deb_etal_05} proposed a kinematic diagnostic of B/P
bulges in face-on ($i<30^{\circ}$) galaxies, namely a double minimum
in the fourth-order Gauss-Hermite moment, $h_4$, along the major-axis
of the bar.  These minima occur because at the location of the B/P
shape, the vertical density distribution of stars becomes broader,
which leads to a double minimum in $z_4$, the fourth-order
Gauss-Hermite moment of the vertical density distribution.  The
kinematic moment $h_4$ is then found to be an excellent proxy for the
unobservable $z_4$.  In contrast, the increase in the vertical
scale-height does not produce any distinct signature of a B/P bulge
and the vertical velocity dispersion, $\sigma_z$, is too strongly
dependent on the radial density variation to provide a useful B/P
bulge diagnostic.  \citet{deb_etal_06} showed that the diagnostic
continues to hold even when gas is present since this sinks to a
radius smaller than that of the B/P bulge.

This new kinematic diagnostic has been confirmed observationally by
\citet{mendezabreu_08}.  Their  main  results  are given here.

\begin{figure*}[!t]
\begin{center}$
\begin{array}{ccc}
\includegraphics[width=4cm, bb=110 360 410 1105]{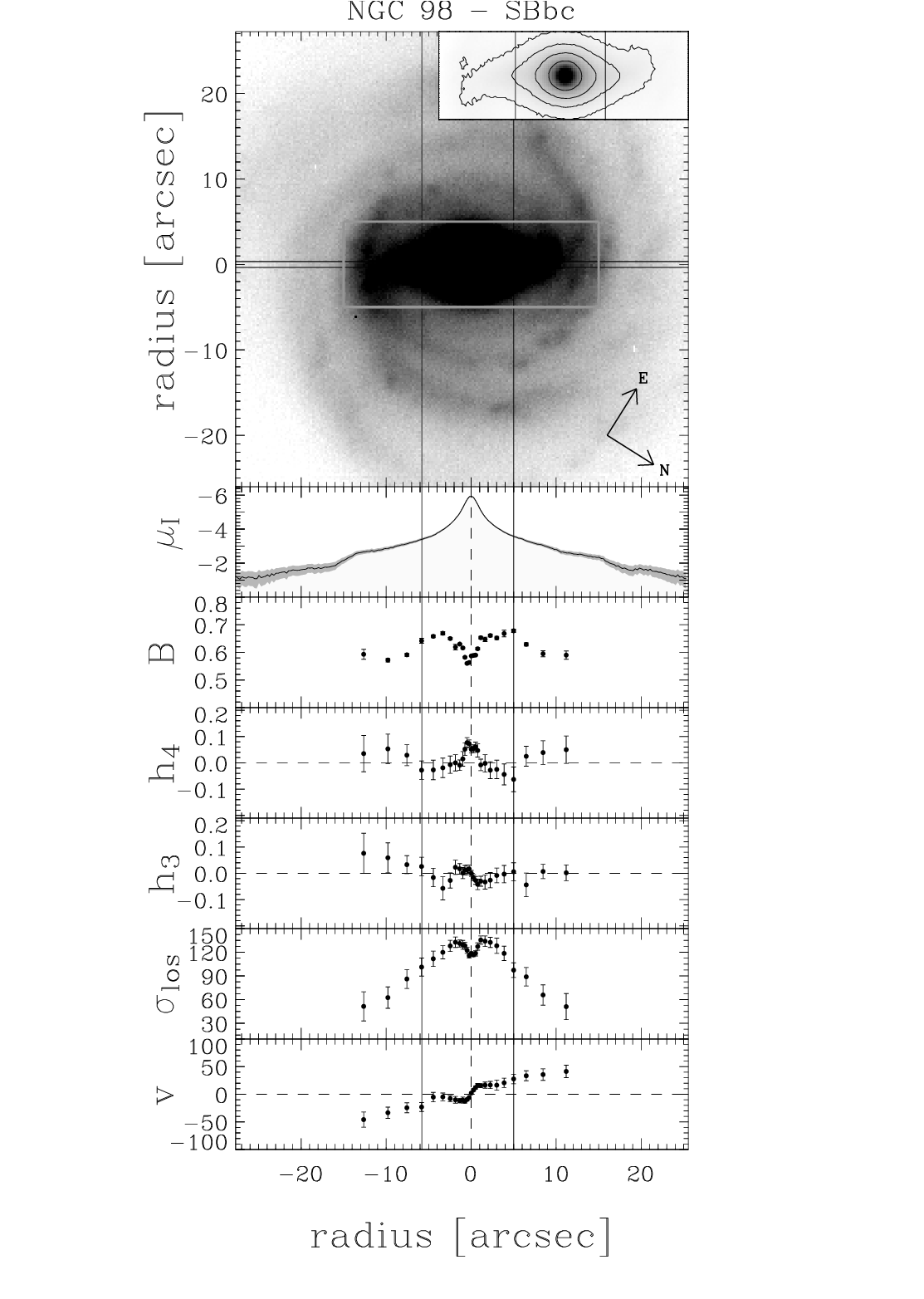} &
\includegraphics[width=4cm, bb=110 360 410 1105]{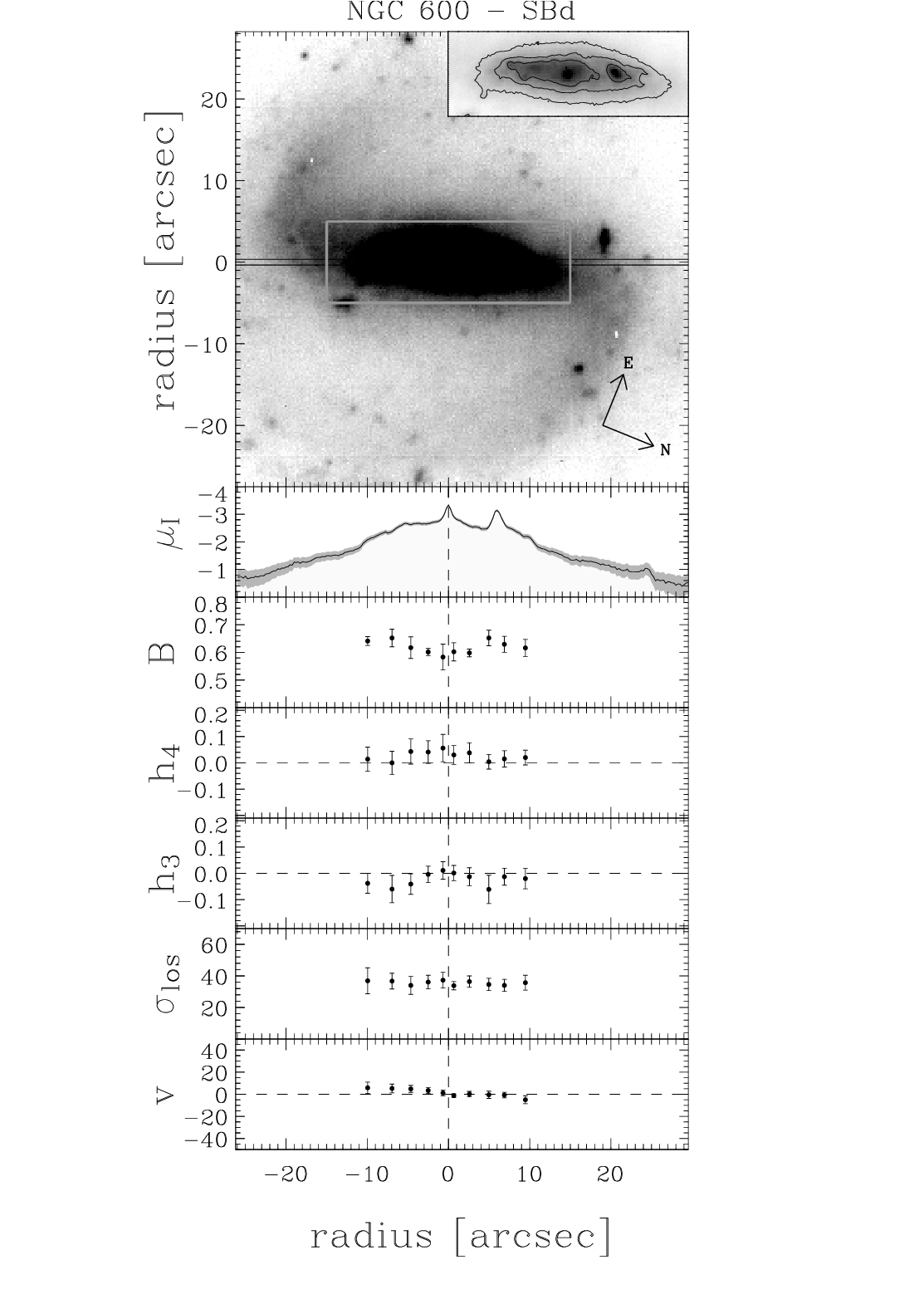} &
\includegraphics[width=4cm, bb=110 360 410 1105]{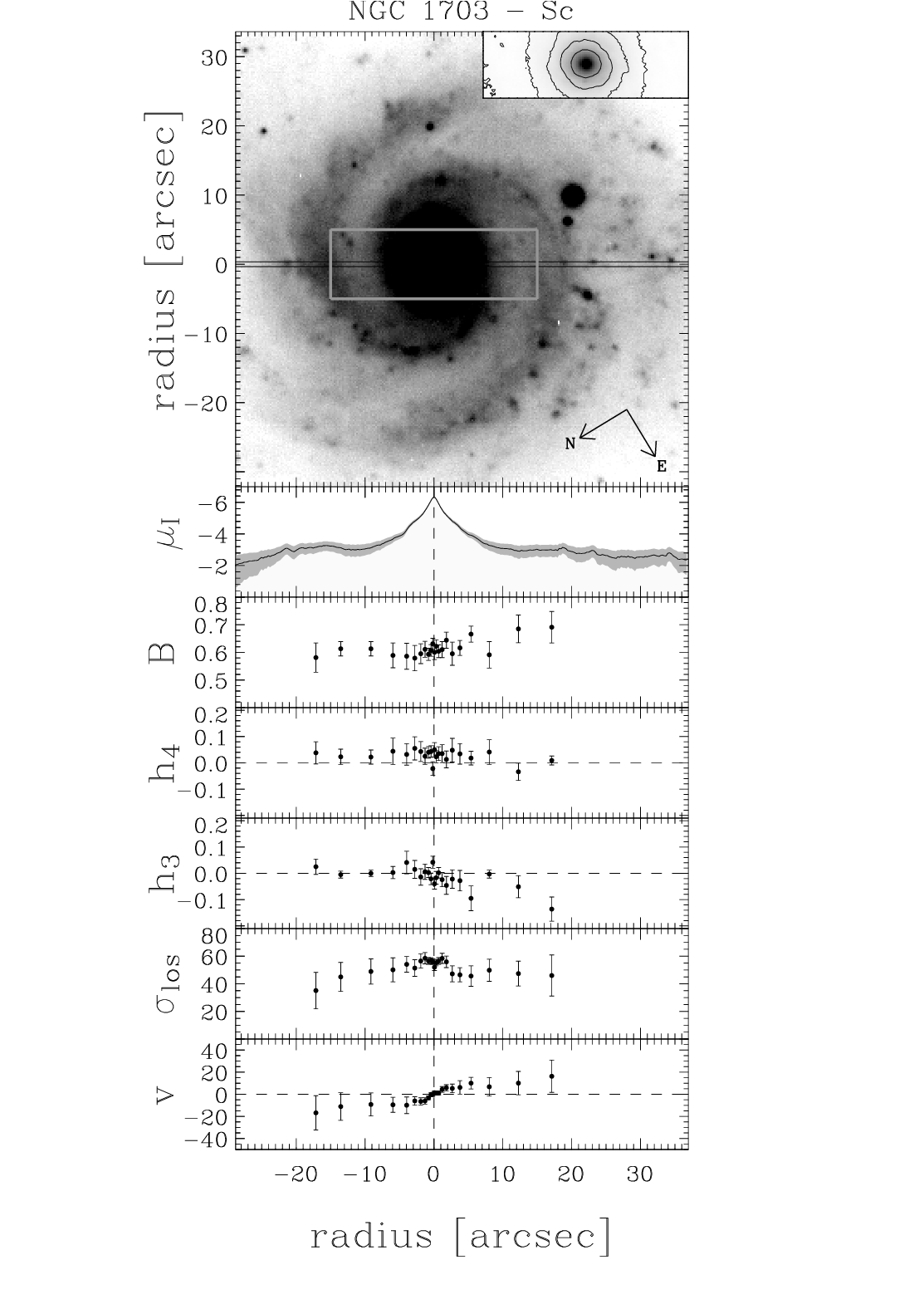}
\end{array}$
\end{center}
\caption{\footnotesize  
  Morphology and stellar kinematics of NGC~98 (left panels), NGC~600
  (central panels), and NGC~1703 (right panels).
  For each galaxy the top panel shows the VLT/FORS2 $R-$band image.
  The slit position and image orientation are indicated.  The inset
  shows the portion of the galaxy image marked with a white box.  The
  gray scale and isophotes were chosen to enhance the features
  observed in the central regions.
  The remaining panels show from top to bottom the profiles of surface
  density, ${\cal B}(0.7,0.9)$, $h_4$, $h_3$, $\sigma$, and mean
  velocity after the subtraction of the systemic velocity, $v$. The
  two vertical lines indicate the location of the $h_4$ minima in
  NGC~98.}
\label{fig:kine}
\end{figure*}

\section{Observations, data-reduction and analysis}

The barred galaxies NGC~98 and NGC~600 were selected as bright and
undisturbed galaxies with a low inclination, a large bar, and modest
extinction in the bar. The nearly face-on unbarred galaxy NGC~1703 was
added as a control object.

The spectroscopic observations were carried out in service mode at the
Very Large Telescope (VLT) at the European Southern Observatory (ESO)
using The Focal Reducer Low Dispersion Spectrograph 2 (FORS2).
The spectra were taken along the bar major axis of NGC~98 and NGC~600
and along the disk major axis of NGC~1703. The total integration time
for each galaxy was 3 hours, in four exposures of 45 minutes each.
Using standard IRAF routines, all the spectra were bias subtracted,
flat-field corrected, cleaned of cosmic rays, corrected for bad
pixels, and wavelength calibrated.  The instrumental resolution was
$1.84\pm0.01$ \AA\ (FWHM) corresponding to $\sigma_{\it inst} = 27$ km
s$^{-1}$ at 8552 \AA .

\subsection{Photometry}

We analyzed the uncalibrated acquisition images from the VLT to derive
the photometric properties of the sample galaxies.
Isophote-fitting with ellipses, after masking foreground stars and bad
pixels, was carried out using the IRAF task ELLIPSE.  Under the
assumption that the outer disks are circular, their inclination was
determined by averaging the outer isophotes.  All 3 galaxies have $i <
30^{\circ}$.
The semi-major axis length, $a_{\rm B}$ of the two bars was measured
from a Fourier decomposition as in \citet{agu_etal_00}.

\subsection{Kinematics}

The stellar kinematics of the three galaxies were measured from the
absorption features present in the wavelength range centered on the
CaII triplet ($\lambda\lambda\,8498,\,8542,\,8662\,$\AA ) using the
Penalized Pixel Fitting method \citep[pPXF,][]{cap_ems_04}.  The
spectra were rebinned along the dispersion direction to a logarithmic
scale, and along the spatial direction to obtain a signal-to-noise
ratio $S/N \ga 40$ \AA$^{-1}$.

A linear combination of the template stellar spectra, convolved with
the line-of-sight velocity distribution (LOSVD) described by a
Gauss-Hermite expansion \citep{gerhar_93, vdm_fra_93} was fitted to
each galaxy spectrum by $\chi^2$ minimization in pixel space. This
allowed us to derive profiles of the mean velocity ($v$), velocity
dispersion ($\sigma$), third ($h_3$), and fourth-order ($h_4$)
Gauss-Hermite moments.  The uncertainties on the kinematic parameters
were estimated by Monte Carlo simulations with photon, read out and
sky noise.
Extensive testing on simulated galaxy spectra was performed to provide
an estimate of the biases of the pPXF method with the adopted
instrumental setup and spectral sampling. The simulated spectra were
obtained by convolving the template spectra with a LOSVD parametrized
as a Gauss-Hermite series and measured as if they were real. No bias
was found in the ranges of $S/N$ and $\sigma$ which characterize the
spectra of the sample galaxies. The values of $h_3$ and $h_4$ measured
for the simulated spectra differ from the intrinsic ones only within
the measured errors \citep[see also][]{ems_etal_04}.

\section{Conclusions}

We have identified for the first time a B/P-shaped bulge in a
face-on galaxy.  In the unbarred galaxy NGC~1703 we had not expected
to find a B/P bulge and we included it in our sample as a control.
The failure to find a double-minimum in $h_4$ is therefore fully
consistent with previous results \citep{chu_bur_04}.  Of the two
barred galaxies, NGC~98 has clear evidence of a B/P bulge while
NGC~600 does not.  The absence of a B/P shape in this galaxy is not
surprising since it appears to not have a bulge.

If we identify the radius of the B/P bulge, $r_{\rm bp}$, with the
location of the minimum in $h_4$, as found in simulations
\citep{deb_etal_05}, then we find $r_{\rm bp} = 0.35$ \mbox{$a_{\rm
    B}$}, where \mbox{$a_{\rm B}$} is the bar semi-major axis.
Similarly, \citet{kor_ken_04} noted that the maximum radius of the
boxy bulge is about one-third of the bar radius.  Simulations also
produce B/P-bulges which are generally smaller than the bar . Since
B/P bulges are supported by resonant orbits, it would be very
instructive to measure this ratio for a sample of barred galaxies.
 

\bibliographystyle{aa}

\end{document}